\documentclass[conference]{IEEEtran}
\IEEEoverridecommandlockouts
\usepackage{hhline}
\usepackage{algorithm}
\usepackage{algpseudocode}



\usepackage{amsmath}
\usepackage{subfig}
\usepackage{enumitem}
\usepackage{multicol}
\usepackage{multirow}
\usepackage{float}
\usepackage{booktabs} 
\usepackage{amssymb}
\usepackage{graphicx}
\usepackage{textcomp}

\def\BibTeX{{\rm B\kern-.05em{\sc i\kern-.025em b}\kern-.08em
    T\kern-.1667em\lower.7ex\hbox{E}\kern-.125emX}}
\begin{document}

\title{A Feature-Driven Approach for Identifying Pathogenic Social Media Accounts\\
}

\author{\IEEEauthorblockN{Hamidreza Alvari, Ghazaleh Beigi, Soumajyoti Sarkar, Scott W. Ruston, \\Steven R. Corman, Hasan Davulcu, Paulo Shakarian}
\textit{Arizona State University}\\
Tempe, USA \\
\{halvari, gbeigi, ssarka18, scott.ruston, steve.corman, hdavulcu, shak\}@asu.edu}

\maketitle

\begin{abstract}
Over the past few years, we have observed different media outlets' attempts to shift public opinion by framing information to support a narrative that facilitate their goals. Malicious users referred to as ``pathogenic social media" (PSM) accounts are more likely to amplify this phenomena by spreading misinformation to viral proportions. Understanding the spread of misinformation from account-level perspective is thus a pressing problem. In this work, we aim to present a feature-driven approach to detect PSM accounts in social media. Inspired by the literature, we set out to assess PSMs from three broad perspectives: (1) user-related information (e.g., user activity, profile characteristics), (2) source-related information (i.e., information linked via URLs shared by users) and (3) content-related information (e.g., tweets characteristics). For the user-related information, we investigate malicious signals using causality analysis (i.e., if user is frequently a cause of viral cascades) and profile characteristics (e.g., number of followers, etc.). For the source-related information, we explore various malicious properties linked to URLs (e.g., URL address, content of the associated website, etc.). Finally, for the content-related information, we examine attributes (e.g., number of hashtags, suspicious hashtags, etc.) from tweets posted by users. Experiments on real-world Twitter data from different countries demonstrate the effectiveness of the proposed approach in identifying PSM users.
\end{abstract}


\section{Introduction}
Recent years have witnessed a surge of manipulation of public opinion and political events by different media outlets and malicious social media actors referred to as ``Pathogenic Social Media" (PSM) accounts~\cite{alvari2018early}. The manipulation of opinion can take many forms from fake news~\cite{shao2017spread} to more subtle ones such as reinforcing specific aspects of text over others~\cite{baron2006persistent}. It has been observed that media aggressively exert bias in the way they report the news to sway their reader's knowledge. On the other hand, PSM accounts are responsible for ``agenda setting'' and massive spread of misinformation~\cite{alvari2019less}. Understanding misinformation from account-level perspective is thus a pressing problem. 

PSM accounts (1) are usually owned by either normal users or automated bots, (2) seek to promote or degrade certain ideas; and (3) can appear in many forms such as terrorist supporters (e.g., ISIS supporters), water armies or fake news writers. Understanding the behavior of PSMs will allow social media to take countermeasures against their propaganda at the early stage and reduce their threat to the public. Early detection of PSMs in social media is crucial as they are likely to be \textit{key} users to forming malicious campaigns~\cite{DBLP:journals/corr/VarolFMF17}. This is a challenging task for several reasons. First, these platforms are primarily based on reports they receive from their own users
\footnote{https://bit.ly/2Dq5i4M} to manually shut down PSMs which is not a timely approach. Despite efforts to suspend these accounts, many of them simply return to social media with different accounts. Second, the available data is often imbalanced and social network structure, which is at the core of many techniques~\cite{conf/icwsm/WengMA14,Kempe2003,beigi2018similar,Zhang:2013:IIN:2444040.2444143}, is not readily available. Third, PSMs often seek to utilize and cultivate large number of online communities of passive supporters to spread as much harmful information as they can. 


\textbf{Present Work.} In this work, we aim to present an automatic feature-driven approach for detecting PSM accounts in social media. Inspired by the literature, we set out to assess PSMs from three broad perspectives: (1) user-related information (e.g., user activity metrics, profile characteristics), (2) source-related information (e.g., information linked via URLs) and (3) content-related information (e.g., tweets characteristics). For the user-related information, we investigate malicious signals using 1) causality analysis (i.e., if user is frequently a cause of viral cascades)~\cite{alvari2018early} and 2) profile characteristics (e.g., number of followers, etc.)~\cite{kudugunta2018deep} aspects of view. For the source-related information, we explore various properties that characterize the type of information being linked to URLs (e.g., URL address, content of the associated website, etc.)~\cite{baly2018predicting,phuong2014gender,entman1993framing,morstatter2018identifying,kincaid1975derivation}. Finally, for the content-related information, we examine attributes from tweets (e.g., number of hashtags, certain hashtags, etc.) posted by users~\cite{kudugunta2018deep}. This paper describes the results of research conducted by Arizona State University’s Global Security Initiative and Center for Strategic Communication.  Research support funding was provided by the US State Department Global Engagement Center. 

Our corpus comprises three different real-world Twitter datasets, from Sweden, Latvia and United Kingdom (UK).  These countries were selected to cover a range of population size and political history (former Soviet republic, neutral, founding member of NATO). In this study, we pose the following research questions and seek answers for them:

\begin{itemize}
    \item[\textbf{RQ1:}] \textit{Does incorporating information from user activities and profile characteristics help in identifying PSM accounts in social media?}
    \item[\textbf{RQ2:}] \textit{What attributes could be exploited from URLs shared by users to determine whether or not they are PSMs?}
    \item[\textbf{RQ3:}] \textit{Could deploying tweet-level information enhance the performance of the PSM detection approach?}
\end{itemize}

To answer \textbf{RQ1}, we investigate whether or not users who make inauthentic information go viral, are more likely to be among PSM users. By exploring \textbf{RQ2}, we figure out which characteristics of URLs and their associated websites are useful in detecting PSM users in social media. By investigating \textbf{RQ3}, we aim to examine if adding a few content-related information on tweet-level could come in handy while identifying PSMs. Our answers to the above questions lead to a feature-driven approach that uses as little as three groups of user, source and content-related attributes to detect PSM accounts. 

\textbf{Key Ideas and Highlights.} To summarize, this paper makes the following main contributions:
\begin{itemize}
	\item We present a feature-driven approach for detecting PSM accounts in social media. More specifically, we assess maliciousness from user-level, source-level and content-level aspects. Our user-related information include signals in causal users (i.e., if user is frequently a cause of viral cascades) along with their profile characteristics (e.g., number of followers, etc.). For the source-related information, we explore different characteristics in URLs that users share and their associated websites (e.g., underlying themes, complexity of content, etc.). For the content-related information, we examine attributes from tweets (e.g., number of hashtags, certain hashtags, etc.) posted by users.

	\item We conduct a suite of experiments on three real-world Twitter datasets from different countries, using several classifiers. Using all of the attributes, we achieve average F1 scores of 0.81, 0.76 and 0.74 for Sweden, Latvian and U.K. datasets, respectively. Our observations suggest the effectiveness of the proposed method in identifying PSM accounts who are more likely to manipulate public opinion in social media. 
\end{itemize}


\section{Related Work}
The explosive growth of the Web has raised numerous
security and privacy issues. Mitigating these concerns has
been studied from several aspects~\cite{Beigi2018HT,alvari2016non,Cao:2014:ULG:2660267.2660269,Cui:2013:COP:2487575.2487639,beigi2014leveraging,broniatowski2018weaponized,beigi2019protecting,alvari2019extremism,beigi2019privacy,beigi2019privacyHT,beigi2019signed}. Our work is related to a number of research directions. Below, we will summarize some of the state-of-the-art methods in each category while highlighting their differences with our work.

\noindent \textbf{Identifying pathogenic social media accounts.} Pathogenic social media (PSM) accounts are believed to be those who are likely to spread malicious messages to viral proportions~\cite{alvari2018early}. For example, the work of~\cite{Causal2017} uses causal inference to detect PSM accounts. Other works of~\cite{alvari2018early,alvaricausal} utilize time-decay causal inference (using sliding-time window) which allows for early detection of PSMs. Also,~\cite{alvari2019less} proposes a semi-supervised causal inference algorithm that achieves reasonable performance using less labeled data by utilizing unlabeled data.

\noindent \textbf{Social Spam/Bot Detection.} 
DARPA organized a Twitter bot challenge to detect ``influence bots''~\cite{7490315}. Among the participants, the work of~\cite{Cao:2014:ULG:2660267.2660269}, used similarity to cluster accounts and uncover groups of malicious users. The work of~\cite{varol2017online} presented a supervised framework for bot detection which uses more than thousands features. In a different attempt, the work of~\cite{ICWSM1715678} studied the problem of spam detection in Wikipedia using different spammers behavioral features. There also exist some studies in the literature that have addressed (1) differences between humans and bots~\cite{chu2012detecting}, (2) different natures of bots~\cite{varol2017online} or (3) differences between bots and human trolls~\cite{broniatowski2018weaponized}. For example the work of~\cite{chu2012detecting} conducted a series of measurements in order to distinguish humans from bots and cyborgs, in term of tweeting behavior, content, and account properties. To do so, they used more than 40 million tweets posted by over 500 K users. Then, they performed analysis and find groups of features that are useful for classifying users into human, bots and cyborgs. They concluded that entropy and certain account properties can be very helpful in differentiating between those accounts. In a different attempt, some other studies have tried to differentiate between several natures of bots. For instance, in the work of~\cite{varol2017online}, authors performed clustering analysis and revealed specific behavioral groups of accounts. Specifically, they identified different types of bots such as \textit{spammers}, \textit{self promoters}, and \textit{accounts that post content from connected applications}, using manual investigation of samples extracted from clusters. Their cluster analysis emphasized that Twitter hosts a variety of users with diverse behaviors; that is in some cases the boundary between human and bot users is not sharp, i.e. some account exhibit characteristics of both.

\noindent \textbf{Fake News Identification.} A growing body of research is addressing the impact of bots in manipulating political discussion, including the 2016 U.S. presidential election~\cite{shao2017spread} and the 2017 French election~\cite{ferrara2017}. For example,~\cite{shao2017spread} analyzes tweets following recent U.S. presidential election and found evidences that bots played key roles in spreading fake news.

\section{Experimental Data}
\begin{figure*}[t]\center
	\includegraphics[width=0.25\textwidth]{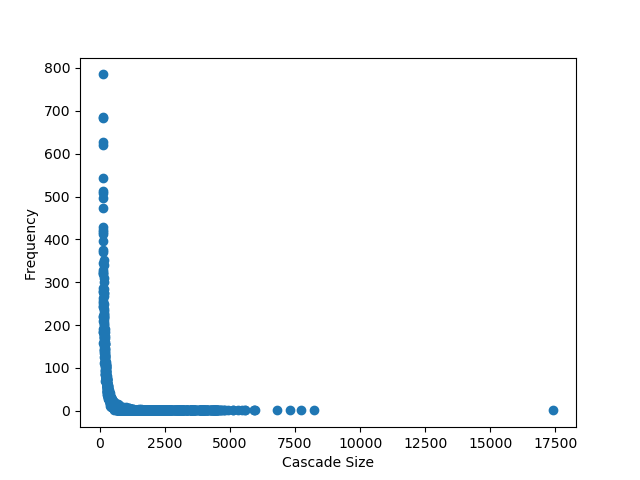}
	\includegraphics[width=0.25\textwidth]{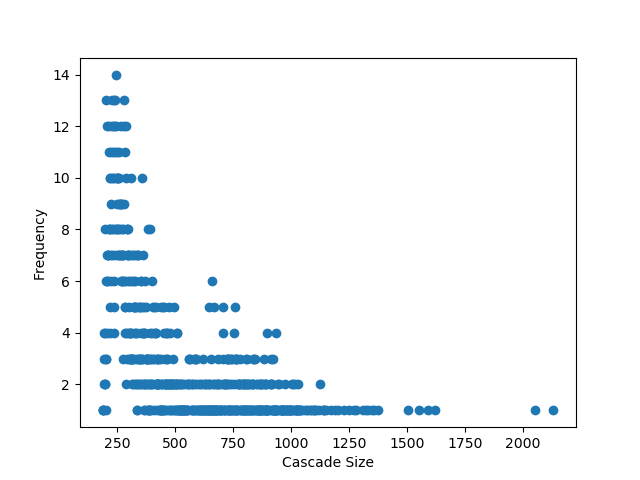}
	\includegraphics[width=0.25\textwidth]{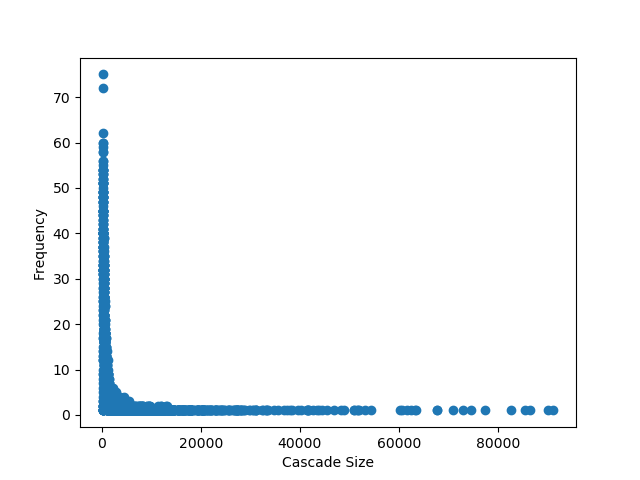}
	\caption{From left to right: Frequency plots of cascade size for Sweden, Latvia and UK datasets.}
	\label{fig:casc}
\end{figure*}
We collect three real-world Twitter datasets with different number of users and tweets/retweets from three countries, Sweden, Latvia and United Kingdom (UK). These countries were selected to cover a range of population size and political history (former Soviet republic, neutral, founding member of NATO). Description of the data is demonstrated in the Table~\ref{tb:st}. We use subsets of datasets from Nov 2017 to Nov 2018. 
Each dataset has different fields including user ID, retweet ID, hashtags, content, posting time as well as user profile information such as Twitter handles, number of followers/followees, description, location, protected, verified, etc. The tweets were collected using a predefined set of keywords and hashtags, and if they were geo-tagged in the country or user profile includes the country. We use subsets of the datasets with different number of cascades of different sizes and duration. 


In our datasets, users may or may not have participated in viral cascades. We chose to use threshold $\theta=20$ and take different number of viral cascades for each dataset with at least 20 tweets/retweets. We depict frequency plots of different cascade size for all datasets in Figure~\ref{fig:casc}. For brevity, we only depict cascades size greater than 100 tweets/retweets. 
\begin{table*}[t]
\small
	\centering
	\caption{Description of the datasets used in this paper.}
	\begin{tabular}{|l|c|c|c|c|c|}
	\cline{1-6}
		\textbf{Dataset} & \textbf{\# Tweets/Retweets} & \multicolumn{2}{c|}{\textbf{\# Labeled Users}} & \textbf{\# Viral Cascades} & \textbf{\# URLs}\\
		\hhline{======}
		& &  \textbf{Suspended} & \textbf{Active} & & \\ \cline{3-4}
		\textbf{Sweden} & 780,250 & 16,010 & 48,030 & 12,174 & 160,702 \\
		\textbf{Latvia} & 323,305 & 10,862 & 32,586 & 1,957 & 76,032 \\
		\textbf{UK} & 254,915 & 4,553 & 13,659 & 21,429 & 41,332 \\
		\cline{1-6}
	\end{tabular}
	\label{tb:st}
\end{table*}


\section{Identifying PSM Users}
\begin{figure}[t]\center
	\includegraphics[width=0.5\textwidth]{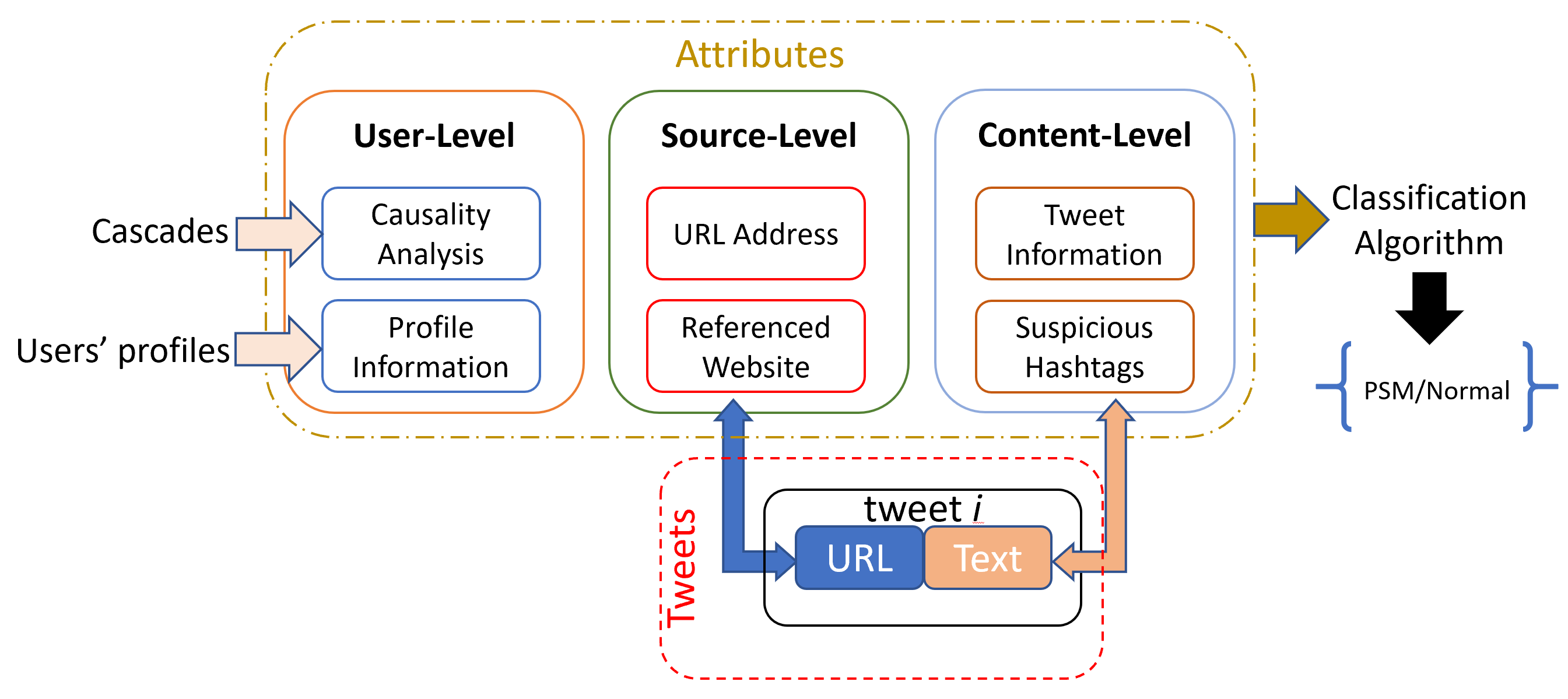}
	\caption{The proposed framework for identifying PSM users. It incorporates three groups of attributes into a classification algorithm.}
	\label{fig:approach}
\end{figure}
In this work, we take a machine learning approach (Figure~\ref{fig:approach}) to answer the research questions posed earlier in the Introduction. More specifically, we incorporate different sets of malicious behavior indicators on user-level, source-level and content-level to detect PSM users. In what follows, we describe each group of the attributes that will be ultimately utilized in a supervised setting to detect PSMs in social media.

\subsection{User-Level Attributes}
We first set out to answer \textbf{RQ1} and understand attributes on the user level that could be exploited in order to identify PSMs in social media. In particular, we investigate two broad categories of attributes: causality-related and profile characteristics. 
\subsubsection{Malicious Signals in Causal Users}
Research has shown that user activity metrics are causally linked to viral cascades to the extent that malicious users who make harmful messages go viral are those with higher causality scores~\cite{alvari2018early}. Accordingly, we set out to investigate if incorporating causality scores in the form of attributes in a machine learning approach, can help identify users with higher malicious behavior in social media. More specifically, We leverage the causal inference introduced in~\cite{alvari2018early} to compute a vector of causality attributes for each user in our dataset. Later, these causal-based attributes will be incorporated to our final vector of attributes that will be fed into a classifier. The causal inference takes as input cascades of tweets/retweets built from the dataset. We follow the convention of~\cite{Goyal:2010} and assume an \textit{action log} $\mathcal{A}$ of the form \textit{Actions(User,Action,Time)}, which contains tuples $(i,a_i,t_i)$ indicating that user $i$ has performed action $a_i$ at time $t_i$. For ease of exposition, we slightly abuse the notation and use the tuple $(i,m,t)$ to indicate that user $i$ has posted (tweeted/retweeted) message $m$ at time $t$. For a given message $m$ we define a \textit{cascade} of actions as $\mathcal{A}_m= \{(i,m',t)\in\mathcal{A}|m'=m\}$. User $i$ is called $m$-participant if there exists $t_i$ such that $(i,m,t_i)\in\mathcal{A}$. Users who have adopted a message in the early stage of its life span are called \textit{key users}~\cite{alvari2018early}.

In this work we adopt the notion of \textit{prima facie causes} which is at the core of Suppes' theory of probabilistic causation~\cite{suppes1970} and utilize the causality metrics that are built on this theory. According to this theory, \textit{a certain event to be recognized as a cause, must occur before the effect and must lead to an increase of the likelihood of observing the effect}. Accordingly, prima facie causal users for a given viral cascade, are key users who help make the cascade go viral. Finally, according to ~\cite{alvari2018early}, we define 4 causal-based attributes for each user and add them to the final representative feature vector for the given user.

\subsubsection{Malicious Signals in Profile Characteristics}
Having defined our causality-based attributes, we now describe our next set of user-based features. Specifically, for each user, we collect account-level features and add them to the final feature vector for that user. We follow the work of~\cite{kudugunta2018deep} and compute the following 10 features from users' profiles: \textit{Statuses Count, Followers Count, Friends Count, Favorites Count, Listed Count, Default Profile, Geo Enables, Profile Uses Background Image, Verified, Protected}. Prior research has shown promising results using this small set of features ~\cite{ferrara2016rise} with far less number of features than the established bot detection approach, namely, Botometer which uses over 1,500 features. Accordingly, we extend the final feature vector representation of each user by adding these 10 features.

\subsection{Source-Level Attributes} Here, we seek an answer to \textbf{RQ2} and examine malicious behavior from the source-level perspective. Previous research has demonstrated the differences between normal and PSM users in terms of their shared URLs~\cite{alvarihawkes}. Accordingly, we take URLs posted by users as source-related information that could be used in our PSM user detection approach. Specifically, we set out to understand several characteristics of each URL from two broad perspectives: (1) URL address and (2) content collected from the website it has referenced.

\subsubsection{URL Address}

\paragraph{\textbf{Far-right and pro-Russian URLs}} Here, we examine if the given URL refers to any of the following far-right websites: \textit{https://voiceofeurope.com/, https://newsvoice.se/, https://nyadagbladet.se/, https://www.friatider.se/} or the pro-russian website \textit{https://ok.ru/}. We further note that each user may have posted multiple URLs posted in our data. To account for that, we compute the average of these attribute values for each user. Ultimately, this list leads to a vector of 5 values for each URL shared by each user in our dataset. We leave examining other malicious websites to future work.

\paragraph{\textbf{Domain Extensions}} Previous research on assessing news articles credibility suggests looking at their URLs~\cite{baly2018predicting} to examine if they contain features such as whether a website contains the \textit{http} or \textit{https} prefixes, or \textit{.gov}, \textit{.co} and \textit{.com} domain extensions. Likewise, we investigate if the URLs in our dataset contain any of these 5 features by counting the number of times each URL triggers one of these attributes and taking the average if user has shared multiple of such URLs. This additional attribute vector will be added to the final attribute vector for each user. 

\subsubsection{Referenced Website Content}
\paragraph{\textbf{Topics}} We further investigate whether or not incorporating the underlying topics or themes learned from the text of the websites, could help us to build a more accurate approach to identify malicious activity. More specifically, we first set out to extract the content from each URL shared by users. To learn the topics, We follow the procedure described in~\cite{phuong2014gender} and train Latent Drichlet Allocation (LDA) \cite{blei2003latent} on the crawled contents of the websites associated with each URL in the training set. 
This way, we obtain a fine-grained categorization of the URLs by identifying their most representative topics as opposed to a coarser-grained approach that uses predefined categories (e.g., sports, etc.). Using LDA also allows for uncovering the hidden thematic structure in the training data. Furthermore, we rely on the document-topic distribution given by the LDA (here each document is seen as a mixture of topics) to distinguish normal users from highly biased users. After training LDA, We treat each new document and measure their association with each of the $K$ topics discovered by LDA. We empirically found $K=25$ to work well in our dataset. Thus, each document is now treated as a vector of 25 probabilistic values provided by LDA's document-topic distribution- this feature space will be added to the final set of the features built so far. Finally, note that for users with more than one URL, we take the average of different probabilistic feature vectors.

\paragraph{\textbf{Has Quote}} 
Social science research has shown that news agencies seek to make a piece of information more noticeable, meaningful, and memorable to the audience~\cite{entman1993framing}. This increases the chance of shifting believes and perceptions. One way to increase salience of a piece of information is emphasizing it by selecting particular facts, concepts and quotes that match the targeted goals~\cite{entman1993framing,scheufele2006framing,dellavigna2007fox}. We thus check the existence of quotes within the referenced website content as an indicator of malicious behavior-- this results in a single binary feature. Each user may post more than one URL. To account for this, We take the average values of this feature for each user. We observe that the PSM users' mean scores for this feature are 0.04 (Sweden), 0.05 (Latvia) and 0.04 (UK). Normal users have mean scores of 0.05 (Sweden), 0.05 (Latvia), and 0.03 (UK). We also deploy two-tailed two-sample t-test with the null hypothesis that value of this feature is not significantly different between normal and PSM accounts. Table.~\ref{tb:complex_read} summarizes the p-values for this test with significance level $\alpha=0.01$. Results show that the null hypothesis could not be rejected. However, we still include this feature to see whether or not it helps in identifying PSMs in practice.

\paragraph{\textbf{Complexity}} Research has shown that complexity of the given text could be different for malicious and normal users~\cite{morstatter2018identifying}. We thus use complexity feature to see whether or not it aids the classifier in finding users who create and share malicious content. We follow the same approach as in~\cite{morstatter2018identifying} and approximate the complexity of reference website content as follows:
\begin{equation}
    \text{complexity} = \frac{\text{number of unique part-of-speech tags}}{\text{number of words in the text}}
\end{equation}

The higher this score is, the more complex the given context is. Surprisingly, our initial analysis show that mean of complexity score of website content by PSMs are 0.53 (Sweden), 0.54 (Latvia) and 0.51 (UK) while mean of complexity score of website contents shared by normal users are 0.46 (Sweden), 0.51 (Latvia), and 0.48 (UK). This shows contents shared by PSMs have higher complexity than those shared by normal users. We also deploy one-tail two-sample t-test with the null hypothesis that content of URLs shared by normal are more complex than those shared by PSMs. Table.~\ref{tb:complex_read} summarizes the p-values showing that the null hypothesis was rejected at significance level $\alpha=0.01$. This indicates that content of websites referenced by PSM users are more complex than those shared by normal users.

\paragraph{\textbf{Readability}} According to~\cite{horne2017just}, readability of a given context can affect engagement of the individuals with the given piece of information. Therefore, readability of the referenced website content is another important feature which could be useful in distinguishing PSMs and normal users. We hypothesize that PSM users may share information with higher readability to increase the chance of transferring the concept and creating malicious content. We use Flesch-Kincaid reading-ease test~\cite{kincaid1975derivation} on the text of the provided URLs. The mean readability scores are 61.16 (Sweden), 62.98 (Latvia), 59.08 (UK) for PSMs and 55.44 (Sweden), 56.79 (Latvia), 55.35 (UK) for other normal users. The higher the score is, the more readable the text is. We also deploy one-tail two-sample t-test with the null hypothesis that content of URLs shared by normal users are more readable than those shared by PSMs. Table.~\ref{tb:complex_read} summarizes the p-values indicating that the null hypothesis was rejected at significance level $\alpha=0.01$.

These results show that the content of URLs shared by PSM accounts are more complex yet more readable than those shared by normal users. Therefore, these two features, complexity and readability, could be a good indicator to distinguish between normal and PSMs.

\begin{table}[t]
\small
	\centering
	\caption{Results of p-values at significance level $\alpha = 0.01$. The null hypotheses for complexity and readability tests are refuted.}
	\begin{tabular}{|l|c|c|c|} 
		\cline{1-4}
		\textbf{Feature} & \textbf{Sweden} & \textbf{Latvia} & \textbf{UK}\\
		\cline{1-4}
		\hhline{====}
		\bf{Has Quote} & 0.29 & 0.36 & 0.32 \\\hline
		\bf{Complexity} & 4.95e-50 & 3.23e-07 & 6.12e-08\\\hline
		\bf{Readability} &  5.56e-27 &  4.2e-03 & 1.9e-14 \\\hline
	\end{tabular}
	\label{tb:complex_read}
\end{table}

\paragraph{\textbf{Unigrams/Bigrams}} We use TF-IDF weighting for extracted word-level unigrams and bigrams. This feature gives us both importance of a term in the given context (i.e., term frequency) and term's importance considering the whole corpus. We remove stop words and select top 20 frequent unigrams/bigrams as the final set of features for this group. Using TF-IDF weighting helps to identify piece of information that is focusing on aspects not emphasized by others. For brevity, we only demonstrate top bigrams in Table~\ref{tb:bigrams}.


\begin{table}[t]\small
	\centering
	\caption{Top selected bigrams for each country.}
	\begin{tabular}{|l|p{6.5cm}|}
		\cline{1-2}
		\textbf{Data}  & \textbf{Bigrams} \\
		\cline{1-2}
        \textbf{Sweden} & asylum seeker, birthright citizenship, court justice, European commission, European Union (EU), European parliament, kill people, migrant caravan, national security, Russian military, school shooting, sexually assault, united nations, white supremacist, police officer \\  \cline{1-2}
        \textbf{Latvia} & Baltic exchange, Baltic security, battlefield revolution, cyber security, depository Estonia, Estonia Latvia, European parliament, European commission, European Union (EU), human rights, Latvian government, nasdaq Baltic, national security, Saeima election, Vladimir Putin \\ \cline{1-2}
        \textbf{UK} & court appeal, cosmic diplomacy, defence police, depression anxiety, diplomacy ambiguity, European Union (EU), human rights, Jewish community, police officer, police federation, political party, rebel medium, sexually liberate, support group, would attacker\\ \cline{1-2}
\end{tabular}
\label{tb:bigrams}
\end{table}

\paragraph{\textbf{Domain Expertise}} The presence of signal words (e.g., specific frames or keywords) could be indicator of existence of malicious behavior in the text. In this work, we hired human coders and trained them based on our codebook\footnote{A codebook is survey research approach to provide a guide for framing categories and coding responses to the the categories definitions.} in order to provide signal words that can help identify suspicious behavior. 
We use the following framing categories: \textit{Anti-immigrant, Crime rampant, Government, Anti-EU/NATO, Russia-ally, Crimea, Discrimination, Fascism}.
For each country and each category, we have a list of corresponding keywords. We have illustrated examples of the keywords used in this study in Table~\ref{tb:keywords}.

  
\begin{table}[t]\small
	\centering
	\caption{Examples of the keywords used in this paper.}
	\begin{tabular}{|l|p{6.5cm}|}
		\cline{1-2}
		\textbf{Data}  & \textbf{Keywords} \\
		\cline{1-2}
        \textbf{Sweden} & no-go zones, violence overwhelmed, police negligence,	Nato obsolete,	bilateral cooperation, blighted areas,	increase reported rapes, close police station, EU hypocrisy, anti-immigrant, fatal shootings, badly Sweden, Nato airstrikes\\ \cline{1-2}

        \textbf{Latvia} & Brussels silent, norms international law, bureaucrats, lack trust EU, based universal principles, Russia borders, anti Nato, purely political, European bureaucrats, silence Brussels Washington, rampant, harsh statements concerning, values Brussels silent\\ \cline{1-2}
		
        \textbf{UK} & Brexit, Theresa May, stop Brexit, hard Brexit, post Brexit, leave, referendum, Brexitshambles\\ \cline{1-2}
\end{tabular}
\label{tb:keywords}
\end{table}

\subsection{Content-Level Attributes}
In this section, we aim to understand \textbf{RQ3} by incorporating a few more attributes from the content-level information that could be used to enhance the performance of the PSM user detection. For the content-level information, we only rely on the tweets posted by each user in our dataset.
\subsubsection{Malicious Signals in Tweet-Level Information} We use the following 6 attributes extracted from each tweet~\cite{kudugunta2018deep}: \textit{retweet count, reply count, favorite count, number of hashtags, number of URLs, number of mentions}. If the user has tweeted more than once, we take the average of these features.
\subsubsection{Malicious Signals in suspicious Hashtags}
We further investigate if the given tweet uses any of the known malicious hashtags identified by our human coders. For Sweden, we use the following list of hashtags: \textit{\#Swedistan, \#Swexit, \#sd (far right group), \#SoldiersofOdin, \#NOGOZones}. For Latvia, we use \textit{\#RussiaCountryFake, \#BrexitChaos, \#BrexitVote, \#Soviet, \#RussiaAttacksUkraine}. For UK data, we use \textit{\#StopBrexit, \#BrexitBetrayal, \#StopBrexitSaveBritain, \#StandUp4Brexit, \#LeaveEU}. Similar to the previous attributes, for the users who have posted more than one tweet with these hashtags, we compute the average of the corresponding values. We leave examining other malicious hashtags to future work.

\subsection{Feature-Driven Approach}
Having described the attributes (Table~\ref{tb:features}) used in this work, we now feed them into a supervised classification algorithm to detect PSM users (Figure~\ref{fig:approach}). 
In more details, we feed the profile information and tweets into the different components of the proposed approach. For the user-related information, we require both of the profile characteristics and tweets. We need tweets to build viral cascades and finally compute causality scores for different users. Each cascade contains tuples $(i,m,t)$ indicating that user $i$ has posted (tweeted/retweeted) the corresponding message $m$ at time $t$. Given the cascades, causality features are computed for each user $i$ based on her activity log in our dataset. For the source-level information, we only need to extract URLs from tweets. These URLs are either directly used to compute attributes or to collect the content from the websites to which they have referenced. For the content-related information, we only need tweets in order to compute the content-level attributes. Finally, for each user, we fuse all attributes into a feature vector representation and feed them into a classifier.


\begin{table*}[t]
\small
 \centering
 \caption{Different groups of features used in this work. Final feature vector representation for each user contains 111 features.}
 \begin{tabular}{|c|c|p{10cm}|c|}
 \hline
 & \multicolumn{1}{c|}{\textbf{Feature}} & \multicolumn{1}{c|}{\textbf{Definition}} & \multicolumn{1}{c|}{\textbf{\# Features}} \\
 \hhline{====}
 \parbox[t]{2mm}{\multirow{6}{*}{\rotatebox[origin=c]{90}{\textbf{User Level}}}} & &&\\
 & Causal-based & Attributes computed using causality based metrics & 4\\
 &  &&\\
 & Profile-based & Profile-related features including: \textit{Statuses Count, Followers Count, Friends Count, Favorites Count, Listed Count, Default Profile, Geo Enables, Profile Uses Background Image, Verified, Protected} & 10\\
 &  && \\ 
  \hline

\parbox[t]{2mm}{\multirow{15}{*}{\rotatebox[origin=c]{90}{\textbf{Source Level}}}} & && \\
 & Websites & Presence of far-right and pro-Russian websites & 5 \\
 & Domains & Existence of \textit{http} or \textit{https} prefixes, or \textit{.gov}, \textit{.co} and \textit{.com} domain extensions & 5 \\
& Topics & Features computed by comparing the listing against the learned topic distribution & 25\\
 
 & Has Quote & Single binary feature that shows whether the content of shared URLs contains quote or not.& 1\\
 
 & Complexity & Complexity of content of shared URLs by users. & 1 \\
 & Readability & Readability of content of shared URLs by users. & 1 \\
 & Unigram & TF-IDF scores of highly frequent word-level unigrams extracted from content of URLs shared by users. & 20 \\
 & Bigram & TF-IDF scores of highly frequent word-level bigrams extracted from content of URLs shared by users. & 20 \\
 & Expertise & Presence of signal keywords provided by our coders & 8 \\
 & && \\
 \hline

 \parbox[t]{2mm}{\multirow{7}{*}{\rotatebox[origin=c]{90}{\textbf{Content Level}}}} & &&\\
  & Tweet-based & Tweet-related features including: \textit{retweet count, reply count, favorite count, number of hashtags, number of URLs, number of mentions} & 6\\
 & && \\
 & && \\
 & Hashtags & Presence of malicious hashtags & 5 \\
 & && \\
  \hline
  \end{tabular}
   \label{tb:features}
 \end{table*}



\section{Experiments}
In this section, we conduct experiments on three real-world Twitter datasets to gauge the effectiveness of the proposed approach. In particular, we compare the results of several classifiers and baseline methods. Note for all methods, we only report results when their best settings are used. 
\begin{itemize}
    \item \textbf{Ensemble Classifiers}
    \begin{itemize}
        \item \textbf{Gradient Boosting Decision Tree (GBDT)} We train a Gradient Boosting Decision Tree classifier using the described features. We set the number of estimators as 200. Learning rate was set to the default value of 0.1.     	
	
	    \item \textbf{Random Forest (RF)} We train a Random Forest classifier using the features described. We use 200 estimators and entropy as the criterion.
	
    	\item \textbf{AdaBoost} We train an AdaBoost classifier using the described features. The number of estimators was set to 200 and we also set the learning rate to 0.01.
    \end{itemize}
	
	\item \textbf{Discriminative Classifiers}
	\begin{itemize}
	    \item \textbf{Logistic Regression (LR)} We train a Logistic Regression using $l2$ penalty. We also set the parameter $C=1$  (the inverse of regularization strength) and tolerance for stopping criteria to 0.01.
	    
	    \item \textbf{Decision Tree (DT)} We train a Decision Tree classifier using the features. We did not tune any specific parameter.
	    
	    \item \textbf{Support Vector Machines (SVM)} We use a linear SVM using the attributes described in the previous section. We set the tolerance for stopping criteria to 0.001 and the penalty parameter $C=1$.
	\end{itemize}
    
    \item \textbf{Generative Classifiers}
    \begin{itemize}
        \item \textbf{Naive Bayes (NB)} We train a Multinomial Naive Bayes which has shown promising results for text classification problems~\cite{manning2008introduction}. We did not tune any specific parameter for this classifier
    \end{itemize}
	
	\item \textbf{Baselines}
	\begin{itemize}
	    \item \textbf{Long Short-Term Memory (LSTM)~\cite{kudugunta2018deep}} The word-level LSTM approach here is similar to the deep neural network models used for sequential word predictions. We adapt the neural network to a sequence classification problem where the inputs are the vector of words in each tweet and the output is the predicted label of the tweet. We first use the word2vec~\cite{mikolov2013distributed} embeddings which are trained jointly with the classification model. We use a single LSTM layer of 50 units on the textual content, followed by the loss layer which computes the cross entropy loss used to optimize the model.N  
	    
	    \item \textbf{Account-Level (AL) + Random Forest ~\cite{kudugunta2018deep}} This approach uses the following features of the user profiles: \textit{Statuses Count, Followers Count, Friends Count, Favorites Count, Listed Count, Default Profile, Geo Enables, Profile Uses Background Image, Verified, Protected}. We chose this method over Botometer~\cite{varol2017online} as it achieved comparable results with far less number of features (\cite{varol2017online} uses over 1,500 features)(see also~\cite{ferrara2016rise}). According to~\cite{kudugunta2018deep}, we report the best results when Random Forest (RF) is used.   	
	
	    \item \textbf{Tweet-Level (TL) + Random Forest ~\cite{kudugunta2018deep}.} Similar to the previous baseline, this method uses only a handful of features extracted from tweets: \textit{retweet count, reply count, favorite count, number of hashtags, number of URLs, number of mentions}. Likewise, we use RF as the classification algorithm.
	\end{itemize}

\end{itemize}

\subsection{Results and Discussion}
		

\begin{table*}[t]
\small
	\centering
	\caption{Performance comparison on different datasets using all features.}
	\begin{tabular}{|l|c|c|c|c|c|c|} 
		\cline{1-7}
		\textbf{Classifier} & \multicolumn{2}{c|}{\textbf{Sweden}} & \multicolumn{2}{c|}{\textbf{Latvia}} & \multicolumn{2}{c|}{\textbf{UK}} \\
		\cline{1-7}
		
		& \textbf{F1-macro} & \textbf{F1-score}
		& \textbf{F1-macro} & \textbf{F1-score}
		& \textbf{F1-macro} & \textbf{F1-score}
		\\
		\hhline{=======}
		\bf{GBDT} & \textbf{0.80} & \textbf{0.81} & \textbf{0.76} & \textbf{0.76} & \textbf{0.73} & \textbf{0.74}\\\hline
		\bf{RF} & 0.79 & 0.79 & 0.75 & 0.75 & 0.70 & 0.71\\\hline
		\bf{AdaBoost} & 0.78 & 0.79 & 0.73 & 0.74 & 0.69 & 0.70\\\hline
		\bf{LR} & 0.75 & 0.75 & 0.74 & 0.74 & 0.71 & 0.72\\\hline
		\bf{DT} & 0.69 & 0.69 & 0.71 & 0.71 & 0.69 & 0.69\\\hline		
		\bf{SVM} & 0.73 & 0.74 & 0.73 & 0.70 & 0.72 & 0.70\\\hline
		\bf{NB} & 0.71 & 0.71 & 0.65 & 0.67 & 0.66 & 0.67 \\\hline
		\bf{LSTM} & 0.60 & 0.62 & 0.58 & 0.65 & 0.36 & 0.43 \\\hline
		\bf{AL (RF)} & 0.64 & 0.64 & 0.63 & 0.64 & 0.64 & 0.65\\\hline
		\bf{TL (RF)} & 0.50 & 0.51 & 0.50 & 0.51 & 0.49 & 0.50 \\\hline
	\end{tabular}
	\label{tb:f1_all}
\end{table*}

All experiments were implemented in Python 2.7x and run on a machine equipped with an Intel(R) Xeon(R) CPU of 3.50 GHz with 200 GB of RAM running Linux. We use tenfold cross-validation follows. We first divide the entire set of training instances into 10 different sets of equal sizes. Each time, we hold one set out for validation. This procedure is performed for all approaches and all datasets for the sake of fair comparison. Finally, we report the average of 10 different runs, using F1-macro and F1-score (only for PSM users) evaluation metrics and all features in Table~\ref{tb:f1_all}.

\subsubsection{Performance Evaluation}
For any approach that requires special tuning of parameters, we conducted grid search to choose the best set of parameters. Also, for LSTM, we preprocess the individual tweets in line with the steps mentioned in \cite{soliman2017aravec}. 
We use word vectors of dimensions 100 and deploy the skip-gram technique for obtaining the word vectors where the input is the target word, while the outputs are the words surrounding the target words. To model the tweet content in a manner that uses it to predict whether an account is biased or not, we used LSTM models \cite{hochreiter1997long}. 
For the LSTM architecture, we use the first 20 words in the tokenized text of each tweet and use padding in situations where the number of tokens in a tweet are less than 20. We use 30 units in the LSTM architecture (many to one). The output of the LSTM layer is fed to a dense layer of 32 units with ReLU activations. We add dropout regularization following this layer to avoid overfitting and the output is then fed to a dense layer which outputs the category of the tweets.

\textbf{Observations. } Overall, we make the following observations:
\begin{itemize}
	\item In general, results from different classifiers compared to the baselines demonstrate the effectiveness of the described attributes in identifying PSM users in social media. Thus, the answers to the research questions \textbf{RQ1}--\textbf{RQ3} are all positive, i.e., we could exploit user, source and content-related attributes for identifying PSM users in social media.
	\item Ensemble classifiers using the described features, outperform all other classifiers and baselines. Amongst the ensemble classifiers, Gradient Boosting Decision Trees classifier achieves the best results in terms of both F1-macro and F1-score metrics.
	\item Amongst the discriminative classifiers, linear Support Vector Machines classifier marginally beats Logistic Regression. Decision Tree classifier achieves the worst results in this category.
	\item Overall, Decision Tree and Naive Bayes classifiers achieve the worst performance among all classifiers.
	\item For LSTM, we achieve slightly poor performance than the logistic regression classifier. One reason behind the poor performance of the classifier is the lack of trained word embeddings suited to our dataset. Also, the poor performance might suggest that the sequential nature of the texts might not be very helpful for the task of PSM users detection.
	\item Overall, results on Sweden data demonstrate better performances achieved using the attributes. One reason behind this might be the size of data and higher number of PSMs in Sweden data compared to others. This could also indicate that PSMs in Latvia and UK data are more sophisticated. 
	
\end{itemize}

\subsubsection{Feature Importance Analysis}
We further conduct feature import analysis to investigate what feature group contributes the most to the performance of the proposed approach. More specifically, we use GBDT and perform different 10-fold cross validations using each feature group. We report the F1-score results in Table~\ref{tb:feat_imp}. According to our observations, we conclude that the most significant and less significant feature groups are \textit{source-related} and \textit{content-related} attributes, respectively. 

\begin{table}[t]
\small
	\centering
	\caption{Feature importance on different datasets.}
	\begin{tabular}{|l|c|c|c|} 
		\cline{1-4}
		\textbf{Feature} & \textbf{Sweden} & \textbf{Latvia} & \textbf{UK} \\
		\hhline{====}
		\bf{User} & 0.65 & 0.61 & 0.59\\\hline
		\bf{Source} & \textbf{0.73} & \textbf{0.70} & \textbf{0.68}\\\hline
		\bf{Content} & 0.45 & 0.43 & 0.40\\\hline
	\end{tabular}
	\label{tb:feat_imp}
\end{table}




\section{Conclusion}
In this work, we presented an automatic feature-driven approach for identifying PSM accounts in social media. In particular, we assess the malicious behavior from three broad perspectives: (1) user, (2) source and (2) content-related information. Experiments on real-word data from three countries demonstrate the effectiveness of the proposed feature-driven approach for detecting PSM accounts in social media.

In future, we would like to replicate the study by examining other attributes such as Granger causality-based attributes. Another potential avenue for future work is to investigate if PSM accounts can form some form of an ecosystem where their activities on social media are linked to their activities on news media. 



\bibliographystyle{IEEEtranS}
\bibliography{mybibfile}

\end{document}